\documentstyle[12pt,epsfig]{article}
\def\beq{\begin{equation}}
\def\eeq{\end{equation}}
\def\bea{\begin{eqnarray}}
\def\eea{\end{eqnarray}}
\def\non{\nonumber}

\begin{document}

\begin{center}
{\Large \bf \sf Spectral singularity and non-Hermitian PT-symmetric extension of $A_{N-1}$ type Calogero model without confining potential
  }

\vspace{1.3cm}

{\sf Bhabani Prasad Mandal \footnote{e-mail address: \ \ bhabani.mandal@gmail.com } and Ananya Ghatak \footnote{e-mail address: \ \ gananya04@gmail.com}}

\bigskip

{\em Department of Physics,\\
Banaras Hindu University,\\
Varanasi-221005, INDIA. \\
}

\bigskip
\bigskip

\noindent {\bf Abstract}

\end{center}

We consider non-Hermitian PT-symmetric deformation of $A_{N-1}$ type Calogero model without confining potential to investigate the possible existence of spectral singularity. By considering the Wronskian between asymptotic incoming and outgoing scattering state wave functions, we found that there exist no spectral singularity in this model. We further explicitly show that the transmission coefficient vanishes and the reflection coefficient becomes unity for all values of the energy in such a momentum dependent non-Hermitian PT-symmetric model.

\medskip
\vspace{1in}
\newpage
\section{Introduction}
Non-Hermitian extension of quantum theories has become a frontier topic of extensive research over the past decade \cite{rev}, mainly because of the following reasons. Fully consistent quantum theories (complete real spectrum, probabilistic interpretation and unitary time evolution ) have been developed for certain categories of non-Hermitian system in a Hilbert space equipped with an appropriate inner product. Secondly due to the huge number of applicability of such non-Hermitian models in quantum optics \cite{opt}, in open quantum systems \cite{oqs}, in quantum field theories \cite{qft}, in quasi-exactly solvable (QES) models \cite{qe} etc. Non-Hermitian but combined parity(P) and time reversal(T) invariant extension of some exactly solvable many-particle quantum mechanical systems \cite{ca, gp} in one space dimension have also been investigated recently \cite{bk} -\cite{br}. In particular PT symmetric non-Hermitian deformation of $ A_{N-1} $ Calogero model (related to $ A_{N-1} $ Lie algebra) and $ B_N $ Calogero model (related to $ B_N $ Lie algebra) are considered by different groups. All these non-Hermitian models exhibit generalized exclusion statistics of Haldane type \cite{ha} and are very attractive due to their various applications in condensed matter physics \cite{rf}. It has been observed that the generalized exclusion and exchange statistic parameter differ from each other in presence of PT symmetric non-Hermitian interaction \cite{bk}-\cite{bnd}. Fring has addressed the important question of whether such extensions are meaningful for all remaining Lie algebra (Coxeter group) and if in addition one can make such extension of the models beyond the rational case to trigonometric, hyperbolic and elliptic models \cite{fr}. They have shown all these deformed rational models are integrable and additional interactions are required to maintain the integrability for deformed non-rational models. Different issues related to solvability and/ or integrability of Calogero-Sutherland (CS) model with PT symmetric non-Hermitian interactions are discussed in Ref.\cite{br}. New QES deformation of CS model has also been studied \cite{br}.

Lack of completeness in a non-Hermitian system is associated with exceptional points (EP) or spectral singularity (SS) when two or more eigenvalues along with corresponding eigenfunctions coalesce \cite{ep, ss, lg}. These are the obstructions to develop fully consistent quantum theories with non-Hermitian Hamiltonians and hence should be investigated in such systems. Different signatures of EP/ SS and their consequences have been studied in great details mainly for single particle systems. To the best of our knowledge these singularities have not yet been investigated in the context of non-Hermitian deformed many-particle systems. The purpose of this present article is to explore the possibility  of existence/ non-existence of spectral singular points in $A_{N-1}$ Calogero model with PT-symmetric non-Hermitian long range interaction. Since we are interested to study SS in such models we would like to restrict our discussion to the PT symmetric non-Hermitian deformed $A_{N-1}$ Calogero model without confining potential as it leads to the scattering states \cite{sca, bnd}. 

By considering the Wronskian between asymptotic incoming and outgoing scattering states we show that these states never become linearly dependent for any values of the energy. It suggests the non-existence of spectral singularity in this model. Our result is further supported by explicit calculation of reflection and transmission coefficient in the presence of PT-invariant long range interacting potential. Transmission coefficient vanishes and reflection coefficient becomes unity in this non-Hermitian case. These coefficients receive no modification due to such a
non-Hermitian PT-symmetric deformation of $A_{N-1}$ Calogero model.

Now we present the plan of the paper. In Sec. 2 of this article we discuss basic aspects $ A_{N-1} $ Calogero model and its deformations. Scattering states of this model are constructed. Nonexistence of SS points in such model is shown explicitly in Sec. 3 and Sec. 4 is kept for summary and discussions.

\section{$A_{N-1}$ Calogero model with non-Hermitian PT-long range interaction}
In this section we briefly outline some of the basic features of $A_{N-1}$ type Calogero model \cite{ca} and review our earlier works \cite{sca, bnd} related to PT-symmetry non-Hermitian momentum dependent deformation of such model.
$ A_{N-1} $ Calogero model containing N particle on a line is described by the Hamiltonian 
\beq
 H = -\frac{1}{2}\sum_{j=1}^N\frac{\partial^2}{\partial
x^2_j}+\frac{g}{2}\sum_{j\neq k}\frac{1}{(x_j-x_k)^2}+\frac{\omega^2}{2}
\sum_{j=1}^N x_j^2.
\label{hh} 
\eeq   
$g$ is the coupling of long-range interaction and $\omega$ is the coupling of harmonic confining interaction. The complete set of bound state energy eigenvalues is given as , 
\beq
E_{n_1,n_2,.....n_N}=\frac{N\omega}{2}\left[1+(N-1)\nu\right]+\omega\sum_{j=1}^N n_j
\eeq
where $ n_j $'s are non-negative integer valued quantum number with $ n_j \leq  n_{j+1} $ and $\nu $ is a real positive parameter related to coupling 
constant $
g=\nu^2-\nu $ . Scattering states with continuous spectrum occur when $ \omega=0 $, i.e. in absence of harmonic confining potential.
The Hamiltonian in Eq.(\ref{hh}) is deformed by adding an extra term $ \delta
\sum_{j\neq k}\frac{1}{(x_j-x_k)}\frac{\partial}{\partial x_j},$
which is non-Hermitian but symmetric under combined $PT$
transformation. For a Hamiltonian containing N particles, the P and T transformations are evidently given by

$ P:\; x_j\;\rightarrow\; -x_j,\; p_j\;\rightarrow -p_j $

$ T:\; x_j\;\rightarrow x_j,\; p_j \;\rightarrow\; -p_j,\;i\rightarrow\; -i $

where $j\in [1,2,....N]$ and $ x_j \ \ (p_j\equiv -i\frac{\partial}{\partial x_j})$  denotes co-ordinate (momenta) operator of the $j$-th particle.
Hence the extended Calogero model which we will be
considering here is described by the Hamiltonian 
\beq
{\cal H}_{ext} = -\frac{1}{2}\sum_{j=1}^N\frac{\partial^2}{\partial
x^2_j}+\frac{g}{2}\sum_{j\neq}\frac{1}{(x_j-x_k)^2}+\delta
\sum_{j\neq k}\frac{1}{(x_j-x_k)}\frac{\partial}{\partial x_j}.
\label{nh} 
\eeq 
The eigenvalue problem for the above Hamiltonian can be solved to obtain scattering 
states within a sector of configuration space corresponding to a definite ordering of
particles like $x_1 \geq x_2 \geq \cdots \geq x_N$.  The zero energy ground state
wavefunction of this model is given by, 
\beq 
\psi_{gr}=\prod_{j<k}
(x_j-x_k)^{\nu^\prime}  
\label {b1} 
\eeq 
where the exponent ${\nu}^\prime$ is related to the coupling constants $g$ and $\delta$ through the relation,
\beq 
g=\nu'^2-{\nu}^\prime\left(1+2\delta\right).   
\label {g} 
\eeq 
For the purpose of obtaining non singular ground state eigenfunction
at the limit $x_i \rightarrow x_j$, $\nu^\prime$ should be a non-negative
exponent. This condition restricts the ranges of coupling constants $g$ and $\delta$ as 
(i) $\delta \geq - \frac{1}{2}, ~0>g\geq-(\delta +\frac{1}{2} )^2,$
and (ii) $g \geq 0$ with arbitrary value of $\delta$. 
The general eigenvalue equation associated the Hamiltonian (\ref {nh}) is given by
\bea 
{\cal H}_{ext} \, \psi \, = \,  p^2 \, \psi ,
\label{gen}
\eea 
where $p$ is a real positive parameter.
It is easy to see that the solutions of this eigenvalue equation
 can be written in the form $ \psi =\psi_{gr}
\tau^\prime (x_1, x_2 \cdots x_N)$, where 
 $\psi_{gr}$ represents the modified ground state eigenfunction(\ref {b1}) and 
$\tau^\prime(x_1,x_2 \cdots x_N)$
 satisfies a differential equation like 
\beq
-\frac{1}{2}\sum_{j=1}^N\frac{\partial^2\tau^\prime}{\partial x_j^2} -
({\nu}^\prime - \delta) \sum_{j \neq
k}\frac{1}{(x_j-x_k)}\frac{\partial \tau^\prime}{\partial x_j} = 
p^2 \tau^\prime \, . 
\label{tau} 
\eeq
Next $\tau^\prime(x_1 , x_2 \cdots x_N) $ is assumed to be factorized as,
\beq
\tau^\prime(x_1 , x_2 \cdots x_N) = P_{k,q}^\prime (x)
\chi^\prime(r), 
\label{taup}
\eeq
where $r$ is the radial variable  defined as $r^2=\frac{1}{N}\sum_{i<j}(x_i-x_j)^2$ and 
$P_{k,q}^\prime(x)$s are translationally invariant, symmetric, k-th
order homogeneous polynomials satisfying the differential
equations 
\beq 
\sum_{j=1}^N \frac{\partial^2 P_{k,q}^\prime(x)}{\partial
x_j^2} + ({\nu}^\prime-\delta)\sum_{j \neq
k}\frac{1}{(x_j-x_k)}\left(\frac{\partial}{\partial x_j} -
\frac{\partial} {\partial x_k}\right) P_{k,q}^\prime(x) = 0\, .
\label{poly} 
\eeq 
The index $q$ in $P_{k,q}^\prime (x)$  can take any integral value ranging from 1 to $g(N,k)$, where $g(N,k)$
is the number of independent polynomials which satisfy Eq. (\ref{poly}) for a given N and k \cite{ca}. These polynomials are translationally invariant and satisfy homogeneity property leading to the relation,
\bea
\sum_{j=1}^N \frac{\partial P^{\prime}_{k,q}(x)}{\partial x_j} = 0 \, ,
~~~~~~~~ \sum_{j=1}^N x_j\frac{\partial P^\prime_{k,q}(x)}{\partial x_j}
= kP^\prime_{k,q}(x) \, . 
\label{prop} 
\eea 
Substituting the factorized form Eq. (\ref {taup}) of $\tau(x_1 ,x_2
\cdots x_N)$ in the differential Eq. (\ref{tau}) and making use of
the properties of $P_{k,q}^\prime(x)$, the equation satisfied by
the 'radial' part of the wavefunction is obtained as, 
\beq
-\frac{\partial^2\chi^\prime(r)}{\partial r^2}
-\frac{1+2b^\prime}{r}\frac{\partial\chi^\prime(r)}{\partial r} =
p^2\chi^\prime(r) 
\label{bb} 
\eeq 
with $ b^\prime = \frac{N-3}{2} +k +
({\nu}^\prime -\delta)\frac{N(N-1)}{2}$.
The solution of Eq. (\ref{bb}) can be expressed through the Bessel 
function: $\chi^\prime
(r) = r^{-b^\prime}J_{b^\prime}(pr)$. Hence the scattering state
eigenfunctions of ${\cal H}_{ext}$ in Eq. (\ref {nh})
 with real positive eigenvalue $p^2$ are obtained as
\beq 
\psi =\prod_{j<k}{(x_j-x_k)}^{\nu^\prime}r^{-b^\prime} J_{b^\prime}(pr) 
P_{k,q}^\prime(x)\, .
\label{sol2} 
\eeq 

A more general eigenfunction of ${\cal H}_{ext}$ is constructed such that 
in the asymptotic limit it can be expressed in terms of an incoming wave ($\psi_+$) 
and an outgoing wave ($\psi_-$). For this purpose 
one has to take appropriate linear superposition of all degenerate
eigenfunctions (with eigenvalue $p^2$) of the form Eq. (\ref{sol2}):
\beq
{\psi}_{gen}=
\prod_{j<k}{(x_j-x_k)}^{{\nu}^\prime}\sum_{k=0}^{\infty}\sum_{q=1}^{g(N,k)}
C^{\prime}_{kq} r^{-b^\prime} J_{b^\prime}(pr) P_{k,q}^\prime(x)\, ,
\label{b20} 
\eeq 
where $C^\prime_{kq}$s are the expansion coefficients and depends on the particle momenta. Each of the momenta $p_i$ is product of a radial part $p$ and an angular part $\alpha_i$ i.e. $p_i=p\alpha_i$ \cite{sca}. By doing 
dimensional analysis, we obtain 
$$
C^\prime_{kq} \, = \,  p^{\frac{(3-N)}{2}+
\frac{N(N-1)\delta}{2}} \, {\tilde C}_{kq}^\prime(\alpha_i) \, ,
$$
where ${\tilde C}_{kq}^\prime(\alpha_i)$ depends only on the angular parts
of the momenta.
Using this explicit expression for $C^\prime_{kq}$ and the 
asymptotic properties of Bessel function at $r
\rightarrow \infty$, we obtain the asymptotic form of 
${\psi}_{gen}$ [Eq. (\ref{b20})] as 
$ {\psi}_{gen} \sim {\psi}_{+} + {\psi}_{-} ,$ 
where 
\beq
{\psi}_{p\pm} = {(2\pi
r)}^{-\frac{1}{2}}p^{({n}^\prime-\frac{1}{2})}
\prod_{j<k}{(x_j-x_k)}^ {{\nu}^\prime}r^{-A^\prime}\sum
_{k=0}^{\infty}\sum_{q=1}^{g(N,k)}{{\tilde
C}^\prime}_{kq}(\alpha_i)
 r^{-k}P_{k,q}^\prime(x)
e^{\pm i(b^\prime + \frac{1}{2})\frac{\pi}{2}\mp ipr}.
\label{asy}
\eeq
In the above expression $A^\prime = b^\prime -k = \frac{N-3}{2}
+(\nu^\prime-\delta)\frac{N(N-1)}{2} $ and $n^\prime =
\frac{3-N}{2} + \frac{N(N-1)\delta}{2}$. By choosing the coefficients ${\tilde C}_{kq}^\prime$ in a proper way, the incoming and outgoing wavefunctions can be written in the form of plane wave like \cite{bnd},
\begin{eqnarray}
\psi_{in} &= & C \exp[{i\sum_{j=1}^{N}p_j x_j}]  \nonumber \\
\psi_{out} &= & C e^{-i\pi{{\nu}^{ \ \prime}\frac{N(N-1)}{2}}}\exp[\,i\sum_{j=1}^N
x_{j} \,p_{N+1-j}\,]  
\label{io}
\end{eqnarray}
where $p_j \leq p_{j+1}$, $p^2=\sum_{j=1}^N p_j^2 $ and $\sum_{j=1}^N p_j = 0$.
These wavefunctions will be used in the next section for further calculations.

\section{Spectral singularity and $A_{N-1}$ Calogero model without confining potential}

In this section we investigate the possible existence of SS in this non-Hermitian many-particle system. If $\psi_{K\pm}(x)$ denotes the solutions for a complex scattering problem in one dimension having continuous positive energy, $H\psi(x)=K^2\psi(x)$,  satisfying the asymptotic boundary conditions, 
 $$\psi_{K\pm}(x)\rightarrow e^{\pm iKx} \mbox{ as } x\rightarrow \pm\infty,$$ 
(i.e. the Jost solutions), then there will be a spectral singularity at $K=K_* $ only when $\psi_{K_*\pm}(x)$ are linearly dependent, at  $K=K_*$ \cite{ss}. This implies that the Wronskian between the two asymptotic solutions $\psi_{K_+}$ and $\psi_{K_-}$ will vanish at spectral singular point $K=K_* $, i.e, 
\beq
W[\psi_{K_+},\psi_{K_-}]=\psi_{K_*+} \psi_{K_*-}^\prime-\psi_{K_*-} \psi_{K_*+}^\prime =0.
\label{wr}
\eeq

We use this well known result to find the possible existence of SS points in N-particle non-Hermitian system. For this purpose we have to construct the general asymptotic wavefunction using Eq. (\ref{io}) in terms of individual momenta $p_j, j=1,2,..N$ as
\beq
\psi_{\pm} = A_{\pm} \exp[{i\sum_{j=1}^{N}p_jx_j}] + B_{\pm} e^{i\pi\phi}\exp[\,i\sum_{j=1}^Nx_{j} \,p_{N+1-j}\,] \ \ \ \ \{ x_j\}\rightarrow \pm \infty 
\label{qw}
\eeq
where $\phi=-{{\nu}^{ \ \prime}\frac{N(N-1)}{2}}$ and $A_{\pm}$, $B_{\pm}$ are possibly $\{p_j\}$-dependent complex coefficients. The individual particle momenta is restricted as,  $p_j \leq p_{j+1}$, $p^2=\sum_{j=1}^N p_j^2 $ and $\sum_{j=1}^N p_j = 0$.

The Jost solutions $\psi_{p_{\pm}}$ in terms of their asymptotic behavior for this system are given as,
\begin{eqnarray}
\psi_{p_{+}}(\{ x_j\}) &\rightarrow &  \exp[{i\sum_{j=1}^{N}p_j x_j}] \ \ \ \ \{ x_j\}\rightarrow  \infty, \nonumber \\
\psi_{p_{-}}(\{ x_j\}) &\rightarrow &  e^{i\pi\phi}\exp[\,i\sum_{j=1}^N
x_{j} \,p_{N+1-j}\,]  \ \ \ \ \{ x_j\}\rightarrow  -\infty 
\label{js}
\end{eqnarray}
From Eqs (\ref{qw}) and (\ref{js}) and with the help of transfer matrix \cite{ss}, 
\begin{eqnarray}
\left(\begin {array}{clcr}
A_{+}  \\
B_{+} \\
\end{array} \right)  & = &  M \left(\begin {array}{clcr}
A_{-}  \\
B_{-} \\\end{array} \right) , \ \ M \ \mbox{is} \ 2\times 2 \ \mbox{matrix}
\end{eqnarray}
we write the Jost solutions $\psi_{p_{+}}(\{x_j\}), \psi_{p_{-}}(\{x_j\})$ for the asymptotic limit of $\{x_j\}\rightarrow  \mp \infty $ as,
\begin{eqnarray}
\psi_{p_{-}}(\{ x_{j}\})&=& M_{12}(\{p_j\})\exp[{i\sum_{j=1}^{N}p_j x_j}]+M_{22}(\{p_j\})e^{i\pi\phi}\exp[\,i\sum_{j=1}^N
x_{j} \,p_{N+1-j}\,] \ \ \ \ \ \ \ \ \{x_j\}\rightarrow  +\infty ,  \nonumber \\
\psi_{p_{+}}(\{x_j\})&=& \frac{M_{22}(\{p_j\})\exp[{i\sum_{j=1}^{N}p_j x_j}]+M_{21}(\{p_j\})e^{i\pi\phi}\exp[\,i\sum_{j=1}^N x_{j} \,p_{N+1-j}\,]}{det M\left(\{p_j\}\right)} \ \ \ \{x_j\}\rightarrow  -\infty  \nonumber \\
\label{js2}
\end{eqnarray}
Now it is straight forward to calculate the Wronskian for $\{x_j\}\rightarrow  +\infty$ using Eqs (\ref{js}) and (\ref{js2}) as,

\begin{eqnarray}
W[\psi_{p_+},\psi_{p_-}]&=&\psi_{p_*+} \psi_{p_*-}^\prime-\psi_{p_*-} \psi_{p_*+}^\prime \nonumber \\
&=& i M_{22}(\{p_j\})\left[p_i-p_{N+1-i}\right] e^{i\pi\phi}\exp[\,i\sum_{j=1}^N x_{j} \,p_{N+1-j}\,]\exp[{i\sum_{j=1}^{N}p_j x_j}] \ \ \nonumber \\
\end{eqnarray}
The prime in the above equation denotes differentiation with respect to $x_i$. Now the Wronskian will vanish if either $M_{22}(\{p_j\})=0$ or 
\beq
p_{i}=p_{N+1-i}.
\label{p}
\eeq
However $M_{22}$ is not equal to zero unless the reflection co-efficient R is infinite \cite{ss}. We have shown explicitly towards the end of this section that in such deformed PT-symmetric many particle system R is always unity. This implies $M_{22}\neq 0$ and hence SS can occur in such systems if Eq. (\ref{p}) is satisfied. However Eq. (\ref{p}) along with the restriction $p_j \leq p_{j+1}$ has only solution $p_j=0$ for all j. 
The non-existence of SS in such non-Hermitian many-particle system now solely depends on the behavior of reflection co-efficient R of the spectrum. The same conclusion can also be obtained by considering the asymptotic wavefunctions [Eq. (\ref{asy})] in terms of radial and angular parts. Now to find the reflection co-efficient R in such systems, for simplicity we start with 2-body scattering i.e. N=2.

The 2-body scattering wavefunction in presence of non-Hermitian interaction can be written compactly as, 
\begin{equation}
\psi_{(2)}=A_2r^cp^{n'-1/2} J_{b'}(pr)
\end{equation}
where, $c=\nu^{\prime}-b'$, $n'=1/2+\delta$ and all the $r$-independent terms are included in $A_2$. The asymptotic behavior of the above wavefunction, 
\beq
\psi_{(2)\pm}= A'_2r^{c-1/2}p^{n'-1/2}e^{\mp ipr}. 
\eeq
We consider the incoming wavefunctions at the region $r<r_-$ as,
\beq
\psi_{(2)\mbox{in}}= r^{c-1/2}p^{n'-1/2}\left(Ae^{- ipr}+Be^{ipr}\right)
\eeq
 and the outgoing wavefunction, 
\beq
\psi_{(2)\mbox{out}}= D r^{c-1/2}p^{n'-1/2}e^{- ipr}
\eeq
is considered at the region $r>r_+$. $r_\pm$ are some reference points where the wavefunctions satisfy the boundary conditions.

We calculate the constants A, B by putting the boundary conditions at $r=r_-$,  $\psi_{(2)\mbox{in}}\mid _{r=r_-}=\psi_{(2)}\mid_{r=r_-}$ and $\psi_{(2)\mbox{in}}^{'}\mid _{r=r_-}=\psi_{(2)}^{'}\mid_{r=r_-}$ as,
\begin{eqnarray}
A=\frac{\left[cr_{-}^{-2}+p^{2}\right]J_{b'}r_{-}^{1/2}-\left[\left(c-1/2\right)r_{-}^{1/2}J_{b'}+r_{-}^{1/2}J_{b'}^{'}\right]\left[cr_{-}^{-1}-ip\right]}{p^{n'-1/2}\left[2p^{2}+2ipcr_{-}^{-1}\right]e^{-ipr_{-}}} \nonumber \\
B=\frac{\left[cr_{-}^{-2}+p^{2}\right]J_{b'}r_{-}^{1/2}-\left[\left(c-1/2\right)r_{-}^{1/2}J_{b'}+r_{-}^{1/2}J_{b'}^{'}\right]\left[cr_{-}^{-1}+ip\right]}{p^{n'-1/2}\left[2p^{2}+2ipcr_{-}^{-1}\right]e^{-ipr_{-}}} \nonumber \\
\end{eqnarray}
The reflection coefficient $R=\frac{\mid B\mid^{2}}{\mid A\mid^{2}}=1$ for all values of $p$.
Similarly we calculate the constant D by satisfying the boundary conditions at $r=r_+$, $\psi_{(2)\mbox{out}}\mid _{r=r_+}=\psi_{(2)}\mid_{r=r_+}, \psi_{(2)\mbox{out}}^{'}\mid _{r=r_+}=\psi_{(2)}^{'}\mid_{r=r_+}$ as,
\begin{equation}
D=\frac{r_{+}J_{b'}^{2}\left(pr_{+}\right)}{p^{n'-1/2}e^{-ipr_{+}}}.
\end{equation}
The transmission coefficient $T=\frac{\mid D\mid^{2}}{\mid A\mid^{2}}$ vanishes for all values of $p$ in the asymptotic limit $r_-\rightarrow \infty$. 

Now we generalize our result for N-particle scattering. We denote the N-body scattering wave function given in [Eq. (\ref{b20})] as,
\beq
\psi_{(N)}=p^{n'-1/2}F(r,\alpha_i )
\eeq
and the incoming scattering wavefunction is written using the asymptotic behavior of the wavefunction given in Eq. (\ref{asy}) as,
\beq
\psi_{(N)\mbox{in}}= p^{n'-1/2}S(r,\alpha_i )\left( A_1 e^{- ipr}+B_1 e^{ipr}\right).
\label{in}
\eeq
$F(r, \alpha_i)$ and $S(r, \alpha_i)$ take care for all the r-dependence and other factors in [Eq. (\ref{b20})] and Eq. (\ref{asy}) respectively. The above wavefunctions satisfy the boundary conditions at some reference point $r_-$, $\psi_{(N)\mbox{in}}\mid _{r=r_-}=\psi_{(N)}\mid_{r=r_-}$ and $\psi_{(N)\mbox{in}}^{'}\mid _{r=r_-}=\psi_{(N)}^{'}\mid_{r=r_-}$. These leads to the values of $A_1$ and $B_1$ as,
\begin{eqnarray}
A_1 &= & \frac{ipF(r_-,\alpha_i )S(r_-,\alpha_i )-F^{'}(r_-,\alpha_i )S(r_-,\alpha_i )+S^{'}(r_-,\alpha_i )F(r_-,\alpha_i )}{p^{n'-1/2}2ipS^2(r_-,\alpha_i )} 
e^{ipr_-}\nonumber \\
B_1 &= & \frac{ipF(r_-,\alpha_i )S(r_-,\alpha_i )+F^{'}(r_-,\alpha_i )S(r_-,\alpha_i )-S^{'}(r_-,\alpha_i )F(r_-,\alpha_i )}{p^{n'-1/2}2ipS^2(r_-,\alpha_i )} 
e^{-ipr_-}\nonumber \\
\end{eqnarray}
Thus the reflection coefficient $R=\frac{\mid B_1\mid^{2}}{\mid A_1\mid^{2}}=1$ for the entire spectrum in all energies. This indicates clearly that $M_{22}$ can never be zero  establishing the non-existence of SS in the spectrum of  this non-Hermitian many particle system.  

We would like to point out here that, the behavior of transmission and reflection coefficients are exactly same as mentioned in the original Calogero's work \cite{ca}. These coefficients receive no modifications due to the addition of PT-symmetric non-Hermitian interaction that suggests the absence of SS points in such many-particle systems.

\section{Summary and Discussions}
It is extremely important to search for spectral singular points and the exceptional points which are obstacles to develop a fully consistent quantum theory with non-Hermitian interactions. There are several different methods to find these spectral singular points. We have studied the possible existence of such singular points in the PT-symmetric non-Hermitian $A_{N-1}$ Calogero model without harmonic confining potential. We have shown explicitly that the asymptotic scattering states never become linearly dependent as the Wronskian between them is not zero. This suggests the non-existence of SS points in this model. Our results are further supported by the fact that reflection coefficient becomes unity in this non-Hermitian model for all energy values. These scattering coefficients remain unaffected even in the presence of such PT-symmetric non-Hermitian interactions. Non-existence of SS points in the $A_{N-1}$ Calogero model with PT invariant non-Hermitian momentum dependent long range interaction indeed favors the work of Fring \cite{fr} where it has been shown that such momentum dependent non-Hermitian deformations of many particle system are in fact integrable classically. It will be interesting to extend this analysis to the non-Hermitian deformation of non-rational Calogero models, where additional interactions are required to maintain the integrability.

\end{document}